\documentclass[final,5p,twocolumn,times,sort&compress]{elsarticle}

\def\preprintdate{IUHET 596, August 2015}

\def\al{\alpha}
\def\be{\beta}
\def\ga{\gamma}
\def\de{\delta}
\def\ep{\epsilon}

\def\et{\eta}
\def\th{\theta}

\def\ka{\kappa}
\def\la{\lambda}

\def\rh{\rho}

\def\ph{\phi}

\def\ps{\psi}

\def\Ga{\Gamma}

\def\cF{{\cal F}}

\def\cL{{\cal L}}
\def\cM{{\cal M}}

\def\fr#1#2{{{#1} \over {#2}}}
\def\half{{\textstyle{1\over 2}}}
\def\quar{{\textstyle{1\over 4}}}

\def\frac#1#2{{\textstyle{{#1}\over {#2}}}}

\def\lsim{\mathrel{\rlap{\lower4pt\hbox{\hskip1pt$\sim$}}
    \raise1pt\hbox{$<$}}}
\def\gsim{\mathrel{\rlap{\lower4pt\hbox{\hskip1pt$\sim$}}
    \raise1pt\hbox{$>$}}}
\def\sqr#1#2{{\vcenter{\vbox{\hrule height.#2pt
         \hbox{\vrule width.#2pt height#1pt \kern#1pt
         \vrule width.#2pt}
         \hrule height.#2pt}}}}

\def\prt{\partial}

\def\etal{{\it et al.}}

\def\pt#1{\phantom{#1}}

\def\ol#1{\overline{#1}}

\def\sb{\overline{s}{}}
\def\cb{\overline{c}{}}
\def\kfb{(\overline{k_F}){}}

\def\sbh{\hat{\sb}{}}
\def\cbh{\hat{\cb}{}}

\def\psb{\overline{\ps}{}}

\newcommand{\beq}{\begin{equation}}
\newcommand{\eeq}{\end{equation}}
\newcommand{\bea}{\begin{eqnarray}}
\newcommand{\eea}{\end{eqnarray}}
\newcommand{\bit}{\begin{itemize}}
\newcommand{\eit}{\end{itemize}}
\newcommand{\rf}[1]{(\ref{#1})}

\def\epd{s^{(d)}}

\def\mn{{\mu\nu}}
\def\ab{{\al\be}}

\def\abgd{{\al\be\ga\de}}
\def\abkl{{\al\be\ka\la}}
\def\gdmn{{\ga\de\mu\nu}}
\def\klmn{{\ka\la\mu\nu}}

\def\C{\v{C}}
\def\cv{\v{C}erenkov}
\def\G{G_N}
\def\Cd{F^w(d)}

\def\kk{\sqrt{16\pi \G}}
\def\kkt{16\pi \G}
\def\pp{|\vec p|}
\def\ml{|\vec l|}
\def\Fd{\cF^w(d)}
\def\Fdw#1{\cF^{#1}(d)}
\def\nw{n_w}
\def\mw{m_w}

\def\re{{\rm Re}~}
\def\im{{\rm Im}~}

\def\cof#1#2{{\ol s}^{(#1)}_{#2}{}}

\begin{document}

\begin{frontmatter}

\title{Constraints on Lorentz violation from 
gravitational \C erenkov radiation}

\author{V.\ Alan Kosteleck\'y$^1$ and Jay D.\ Tasson$^2$}

\address{$^1$Physics Department, Indiana University,
Bloomington, IN 47405, U.S.A.\\
$^2$Physics Department, St.\ Olaf College, 
Northfield, MN 55901, U.S.A.}

\address{}
\address{\rm 
\preprintdate;
accepted for publication in Physics Letters B}

\begin{abstract}

Limits on gravitational \C erenkov radiation by cosmic rays 
are obtained and used to constrain coefficients for Lorentz violation 
in the gravity sector
associated with operators of even mass dimensions,
including orientation-dependent effects.
We use existing data from cosmic-ray telescopes
to obtain conservative two-sided constraints 
on 80 distinct Lorentz-violating operators of dimensions four, six, and eight,
along with conservative one-sided constraints on three others.
Existing limits on the nine minimal operators at dimension four 
are improved by factors of up to a billion,
while 74 of our explicit limits represent stringent first constraints
on nonminimal operators.
Prospects are discussed for future analyses incorporating
effects of Lorentz violation in the matter sector,
the role of gravitational \C erenkov radiation by high-energy photons,
data from gravitational-wave observatories,
the tired-light effect,
and electromagnetic \C erenkov radiation by gravitons. 

\end{abstract}

\end{frontmatter}

\bigskip
\noindent
{\it 1.\ Introduction.}
A century after its formulation,
General Relativity (GR) is established as a remarkably successful 
classical field theory of gravity.
Extending GR into the quantum domain
is widely believed to require modifications
of one or more of its founding principles, 
and identifying experimental tests to confirm this
forms an interesting challenge.
One central component of GR
is local Lorentz invariance,
which is symmetry under local rotations and boosts.
Experimental tests of this invariance 
have undergone a renaissance in recent years 
\cite{tables},
following the realization 
that tiny observable violations of Lorentz invariance
could arise naturally in some quantum theories of gravity
such as strings
\cite{ksp}.
While impressive sensitivities to a broad range 
of possible violations
in the matter sector have been achieved,
searches for Lorentz violation in the gravity sector 
have been less extensive.
In the present work,
we obtain tight constraints on a large class
of potential Lorentz-violating operators in the pure-gravity sector.

Deviations from local Lorentz invariance in gravity
can be studied using effective field theory,
which offers a model-independent approach
to describing Lorentz-violating effects 
arising in an underlying theory of quantum gravity 
\cite{akgrav}.
Within this approach,
the Lagrange density describing general Lorentz violation for pure gravity 
is a subset of the gravitational Standard-Model Extension (SME)
consisting of the usual Einstein-Hilbert and cosmological-constant terms,
along with a series of additional terms 
containing all possible Lorentz-violating operators.
In each term,
the Lorentz-violating operator is formed from gravitational-field variables
and is contracted with a coefficient
controlling the magnitude of the effects.
The Lorentz-violating operators can be organized in a series
according to increasing mass dimension $d$ in natural units,
with the corresponding coefficients having mass dimensions $4-d$. 
Within the pure-gravity sector of this framework,
most experimental studies
\cite{2007Battat,2007MullerInterf,2009Chung,%
2010Bennett,2012Iorio,2013Bailey,2014Shao,he15}
and theoretical investigations 
\cite{bk,se09,al10,bt11,jt12,yb15}
have focused on minimal operators for Lorentz violation,
which have mass dimension $d=4$.
Some observational consequences of nonminimal operators 
of dimensions $d=5,6$ are known
\cite{bkx},
and experimental studies of nonrelativistic effects of $d=6$ operators 
on short-range gravity have recently been performed
\cite{lo15,hust15,hustiu}.
For reviews see,
for example,
Refs.\ \cite{jt,cw,rb}. 
Here,
we obtain stringent conservative constraints 
on certain Lorentz-violating operators of even dimensions $d\geq 4$,
following from the observation of high-energy cosmic rays
and the consequent limits on vacuum gravitational \cv\ radiation.
 
Electromagnetic \cv\ radiation in ponderable media
has been extensively studied since its discovery
in the early 1930s
\cite{cvexp,cvth}.
It arises when the velocity of a massive charged particle
exceeds the phase velocity of light in a medium,
thereby rendering the particle unstable to radiation of \cv\ light.
In the presence of Lorentz violation,
the vacuum acts like a refractive medium for particles 
with properties controlled by the coefficients for Lorentz violation
\cite{sme}.
Under these circumstances,
a particle travelling in a vacuum 
with velocity exceeding that of light
can produce vacuum \cv\ radiation,
which continues until the particle loses enough energy
to drop below light speed.
The observation of high-energy particles of various species
limits the existence of vacuum \cv\ radiation
and therefore places constraints 
on certain coefficients for Lorentz violation in the matter sector
\cite{co99,st01,ja03,ga04,lp04,al07,al07-2,al08,kl08,ms13,ho09,km12,km13}.
Any single coefficient constraint is normally one-sided
because \cv\ radiation 
is possible only for superluminal particles,
which typically occurs for only one coefficient sign. 

Gravitational \cv\ radiation is an analogous effect
that is hypothesized to occur
when the velocity of a particle exceeds the phase velocity of gravity.
In principle,
this could occur in conventional GR
in the presence of dust, gas, or other media,
but the radiation rate is suppressed by two powers
of the Newton gravitational constant $\G$
and hence is negligible for practical purposes
\cite{pe74,sz71,po72,ch73,pa94}.
However,
in the presence of Lorentz violation,
vacuum gravitational \cv\ radiation 
suppressed by only one power of $\G$ can arise 
and would produce energy losses of particles
travelling over astrophysical distances
\cite{mo01,el05,ki12,la12}.
The observation of high-energy cosmic rays
therefore constrains certain coefficients for Lorentz violation
in the gravity sector.
In this work,
we use observations of the energies and celestial positions 
of cosmic-ray events
to obtain conservative limits on a large class of coefficients 
in the pure-gravity sector,
setting stringent first constraints on many coefficients
and improving certain existing limits by factors of up to a billion.

Observable effects on photon propagation 
arising from Lorentz violation involving operators of arbitrary $d$ 
can be classified in terms of 
anisotropy, dispersion, birefringence,
and whether they affect vacuum propagation
\cite{km09}.
A comparable analysis of quadratic gravitational operators of arbitrary $d$
reveals that a similar classification holds also 
in the gravity sector
\cite{km15}.
We focus here on nonbirefringent vacuum effects 
involving gravitational operators of arbitrary mass dimension,
which can intuitively be viewed as certain components 
of a derivative-dependent effective metric $\sbh^\ab$.
We obtain wave solutions
for this class of modifications to the Einstein field equations,
derive expressions for the rates of vacuum gravitational \cv\ radiation 
of scalars, fermions, and photons,
and apply the results to extract explicit conservative constraints 
on coefficients for Lorentz violation
for even mass dimensions $4 \leq d \leq 8$.
Throughout this work,
we use the notations and conventions of Ref.\ \cite{akgrav}.

\bigskip
\noindent
{\it 2.\ Lorentz-violating gravitational waves.}
The effective gravitational field theory 
containing Lorentz-violating operators of arbitrary mass dimensions 
\cite{akgrav}
can be linearized to produce modified Einstein equations
relevant for weak-field gravity at leading order
in coefficients for Lorentz violation
\cite{bk,bkx,km15}.
Our focus here is on perturbative modifications
that can be written in the form
\beq
G_\mn = 8\pi \G (T_M)_\mn +\sbh^{\al\be} \widetilde R_{\al\mu\be\nu},
\label{eeq}
\eeq
where $G_N$ is the Newton gravitational constant,
$(T_M)_\mn$ is the matter energy-momentum tensor,
$\widetilde R_\abgd \equiv \ep_\abkl \ep_\gdmn R^\klmn/4$
is the double dual of the Riemann tensor,
and $G_{\mn}$
is the Einstein tensor.
All expressions are understood to be linearized
in a flat-spacetime background with Minkowski metric,
$g_\mn = \et_\mn+ h_\mn$.
The operator $\sbh_{\mn}\equiv\sbh_{\nu\mu}$
is formed as a sum of terms containing coefficients 
$(\sb^{(d)})_{\mn}{}^{\al_1 \ldots \al_{d-4}}$ 
for Lorentz violation 
and even powers of derivatives, 
\beq
\sbh_\mn \equiv 
\sum_d 
(\sb^{(d)})_{\mn}{}^{\al_1 \ldots \al_{d-4}} 
\prt_{\al_1} \ldots \prt_{\al_{d-4}},
\label{sdef}
\eeq
with the sum understood to range over even values $d\geq 4$.
The coefficients $(\sb^{(d)})_{\mn}{}^{\al_1 \ldots \al_{d-4}}$
are constant and assumed to be small.
The $d=4$ coefficient $\sb_\mn\equiv (\sb^{(4)})_{\mn}$ 
appears in the minimal Lorentz-violating extension of GR
and has been the subject of various experimental tests
\cite{2007Battat,2007MullerInterf,2009Chung,%
2010Bennett,2012Iorio,2013Bailey,2014Shao,he15}.
Nonrelativistic effects from 
some components of the second term $(\sb^{(6)})_{\mn}{}^{\ab}$ 
have recently been experimentally constrained as well
\cite{lo15,hust15,hustiu}.

The perturbative change to the field equations \rf{eeq} 
preserves diffeomorphism symmetry
even though the background coefficients 
$(\sb^{(d)})_{\mn}{}^{\al_1 \ldots \al_{d-4}}$
violate it.
This can be understood as a consequence
of the spontaneous breaking of diffeomorphism and Lorentz symmetry
\cite{bk05}. 
As a result,
the usual counting of degrees of freedom 
in the metric fluctuation $h_\mn$ holds,
with four auxiliary components, four gauge components,
and two physical gravitational modes relevant to observable physics.
The additional modes arising from the higher derivative powers
appear only at high energies that lie 
beyond the domain of validity of effective field theory.
Note that the form of Eq.\ \rf{eeq} reveals that
$\sbh^\mn$ plays the role of a derivative-dependent shift 
of the metric,
$\et_\mn \to \et_\mn - \sbh_\mn$.
The perturbation also includes Lorentz-invariant contributions,
with effects governed by the trace pieces of the coefficients
$(\sb^{(d)})_{\mn}{}^{\al_1 \ldots \al_{d-4}}$.
More general modifications to the Einstein equations
can be countenanced and classified 
\cite{akgrav,bkx,km15},
but exploring the implications of these lies outside our present scope. 
We remark in passing that odd powers of derivatives in the expression \rf{sdef}
are excluded by the requirements of hermiticity and diffeomorphism invariance.

We seek solutions to the modified Einstein equations \rf{eeq}
representing perturbations of the conventional linearized gravitational waves 
propagating in the Minkowski vacuum.
The wave solutions can readily be found at leading order in $\sbh^\mn$.
The conventional Einstein equations in vacuum 
\beq
R_\mn = 0
\label{zerothd}
\eeq
hold at zeroth order,
which implies that the modified Einstein equations at first order
can be written in the form
\beq
(\et^\ab + \sbh^\ab) R_{\al\mu\be\nu} = 0.
\label{firstd}
\eeq
Note that working at first order in $\sbh^\mn$ in this equation
requires keeping both zeroth- and first-order contributions 
to the contraction of the Riemann tensor with $\et^\ab$,
but only zeroth-order contributions to that with $\sbh^\ab$.
To fix the gauge,
we adopt the modified Hilbert condition
\beq
\prt_\al (\et^\ab + \sbh^\ab) h_{\be\mu} = 
\half \prt_\mu (\et^\ab + \sbh^\ab) h_\ab
\label{hgauged}
\eeq
and the traceless condition $(\et^\ab + \sbh^\ab) h_{\ab} =0$ 
as natural choices,
in light of the interpretation of the perturbation
as a shift of the inverse metric.

To find the wave solutions, 
it is convenient to convert to momentum space
via the ansatz
\beq
h_\mn (x) = A_\mn (l) e^{i l_\al x^\al},
\label{ansatz}
\eeq
where $l^\al$ is the 4-momentum of the gravitational wave
and where as usual only the real part of the right-hand side is taken.
This implies the replacement $\prt_\al \to i l_\al$
can be adopted in the definition \rf{sdef}
whenever $\sbh_\mn$ acts on $h_\mn$.
For $d=4$,
the quadratic momentum dependence implies that 
the result tracks the conventional case 
modulo the deformation of the Minkowski metric.
However,
for $d\geq 6$ the corrections to the usual solutions
involve higher powers of the 4-momentum $l^\al$.

Using the ansatz \rf{ansatz},
the modified Einstein equation \rf{firstd} in the gauge \rf{hgauged}
takes the form
\beq
(\et^\ab + \sbh^\ab) l_\al l_\be A_\mn =0.
\eeq
The resulting dispersion relation for the gravitational waves,
\beq
l_0^2 = \vec l^2 + \sbh^\ab l_\al l_\be,
\label{disp}
\eeq
suggests introducing an effective vacuum refractive index 
$n = n(\vec l)$
for gravitational waves,
given by
\beq
n^2 = 1 - \sbh^\ab \hat{l}_\al \hat{l}_\be,
\label{efri}
\eeq
where $\hat{l}_\al \equiv l_\al/l_0$.
We emphasize that for $d\geq 6$ the refractive index $n$
depends on the momentum and hence the energy of the gravitational wave
and that for all $d$ it receives direction-dependent 
Lorentz-violating contributions 
as well as both isotropic Lorentz-violating ones
and Lorentz-invariant ones.
The group velocity $\vec v_g$ can be obtained by
differentiating $l_0$ with respect to $\vec l$,
yielding
\bea
\hskip -20pt
|\vec v_g| = 1 
+ \half 
\sum_{d} (-1)^{d/2} (d-3)l^{d-4}
\hat l^\mu \hat l^\nu 
\hat l_{\al_1} \ldots \hat l_{\al_{d-4}}
(\sb^{(d)})_\mn{}^{\al_1 \ldots \al_{d-4}}. 
\hskip -20pt
\nonumber\\
\eea

At leading order in $\sbh^\mn$,
the wave vector $l_\al$ can be written
in terms of the conventional wave vector $l^{(0)}_\al$
in the form
\beq
l_\al = l^{(0)}_\al - \half \sbh_\al{}^\be l^{(0)}_\be,
\eeq
where $\sbh_\al{}^\be = \sbh_\al{}^\be (l)$
is evaluated using $l^{(0)}$.
The amplitude of the wave \rf{ansatz}
can similarly be written in terms of the conventional 
plus and cross modes of GR.
To obtain an explicit expression,
it is convenient to work with trace-reversed quantities
defined at first order by 
\beq
\ol h_\mn = h_\mn + \half \et^{(1)}_\mn h_\ab \et^{(1) \ab},
\eeq
where $\et^{(1)}_\mn \equiv \et_\mn - \sbh_\mn$
and $\et^{(1) \mn} \equiv \et^\mn + \sbh^\mn$.
In terms of the trace-reversed amplitude $\ol A_\mn$,
the gauge condition \rf{hgauged} then takes the form
\beq
l_\al (\et^\ab + \sbh^\ab) \ol A_{\be\mu} = 0.
\label{tgauged}
\eeq
This yields an expression for
the trace-reversed amplitude $\ol A_\mn$
in terms of the conventional trace-reversed amplitude $\ol A{}_\mn^{(0)}$,
\beq
\ol A_\mn = 
\ol A{}^{(0)}_\mn - \sbh_{(\mu}{}^\al \ol A{}^{(0)}_{\al\nu)}.
\label{amplituded}
\eeq
where the symmetrization on the indices $\mu$,$ \nu$ 
includes a factor of 1/2.
The result also shows that the graviton polarization matrix $\ep_\mn$
appearing in the matrix elements for quantum processes with gravitons
takes the form
\beq
\ep_\mn = 
N ( \ep^{(0)}_\mn - \sbh_{(\mu}{}^\al \ol \ep^{(0)}_{\al\nu)}),
\label{gravpol}
\eeq
where 
$\ep^{(0)}_\mn$ is the usual graviton polarization matrix
and $N$ is a normalization factor.

\bigskip
\noindent
{\it 3.\ Gravitational \cv\ radiation.}
A particle of any species travelling faster
than the phase velocity of gravity 
becomes unstable to gravitational \cv\ radiation.
The differential rate 
to radiate a single graviton of momentum $l_\mu$
is given by
\beq
\hskip -20pt
d \Ga = \fr{1}{2 \sqrt{\mw^2 + \vec p^2}}
\fr{d^3k}{(2\pi)^32k_0}
\fr{d^3l}{(2\pi)^32l_0}
|\cM|^2 (2 \pi)^4 \de^4(p-k-l),
\quad
\label{diffrate}
\eeq
where $p_\mu$ is the incoming particle momentum
obeying the dispersion relation $p_0^2 = \vec p^2 + \mw^2$
for a particle of species $w$ and mass $\mw$,
$k_\mu = p_\mu - l_\mu$ is the outgoing particle momentum,
and $\cM$ is the matrix element for the decay
in the quantum field theory.
The integrated rate of energy loss is therefore given by
\bea
\hskip -10pt
\fr {dE}{dt} = 
- \fr{1} {4p_0} \int 
\fr{d^3k}{(2\pi)^3 2k_0}\fr{d^3l}{(2\pi)^3}
|\cM|^2 
(2 \pi)^4 \de^4(p-k-l).
\hskip -20pt
\label{enrate}
\eea

Consider first the kinematical aspects 
of the rate of energy loss \rf{enrate}.
Since the applications to follow involve $m_w\ll p_0$,
we can neglect contributions of order $m_w$ times Lorentz violation,
which implies that in the delta function
both $\hat l$ and $\hat k$ are aligned with $\hat p$ at leading order.
Performing the integrals over the outgoing momenta $\vec k$ 
and manipulating the remaining delta function yields
\beq
\hskip -10pt
\fr {dE}{dt} = 
-\fr{1} {8\pp \sqrt{\mw^2 + \vec p^2}}
\int \fr{d^3l}{ (2 \pi)^2 \ml} |\cM|^2
\de ( \cos \th - \cos \th_C ),
\label{rated}
\eeq
where $\th$ is the angle between $\vec p$ and $\vec l$.
The generalized \cv\ angle $\th_C$ can be written as
\beq
\cos \th_C = 
\fr {\sqrt{\mw^2 + \vec p^2}} {\pp} 
\fr 1 {n(\ml)}
+ \fr{\ml}{2\pp} \left(1-\fr{1}{[n(\ml)]^2}\right), 
\label{ca}
\eeq
where the convenient notation $n(\ml) \equiv n(\ml\hat p)$ is used.
The result \rf{ca}
is a function of both $\pp$ and $\ml$,
and it encodes the Lorentz violation 
via the refractive index.

The delta function in the integrand of the rate \rf{rated}
governs physical properties of the gravitational \cv\ radiation.
For example,
in the limit $\ml\ll \pp$
the integrand acquires contributions 
only for radiation at the special \cv\ angle $\th =\th_C$
given by $\cos\th_C = 1/n\be$,
where $\be$ is the particle speed.
This matches the well-known result
for conventional electromagnetic \cv\ radiation in a medium
of refractive index $n$. 
The existence of a possible threshold velocity $\be_{\rm th}$
below which no radiation occurs can also be seen.
For example,
in the above limit $\be_{\rm th} = 1/n$,
which again reproduces the classical result for photon \cv\ radiation.
The maximum angle of emission $\th_{C,\rm ~max} = \cos^{-1}(1/n)$
occurs for an ultrarelativistic particle with $\be\to 1$ in this case.

For calculational purposes,
it is convenient to express the correction to the refractive index 
arising from operators of dimension $d$ as
\beq
n \approx 1 - \half \sum_{d} (-1)^{d/2} \epd \ml^{d-4},
\label{nepd}
\eeq
where $\epd$ is a direction-dependent combination 
of coefficients for Lorentz violation given by
\beq
\epd (\hat l) \equiv 
(\sb^{(d)})^{\mn\al_1 \ldots \al_{d-4}}
\hat l_\mu \hat l_\nu 
\hat l_{\al_1} \ldots \hat l_{\al_{d-4}}
\label{refinds}
\eeq
and having mass dimension $4-d$.
For the special case of an incoming photon
or ultrarelativistic particle with $\be\to 1$,
we can use the form \rf{nepd} for $n$ 
to obtain expressions for the \cv\ angle $\th_C$
valid to leading order in Lorentz violation,
\bea
\cos \th_C &\approx& 
1 + \half \sum_{d} (-1)^{d/2} 
\epd \ml^{d-4}
\left( 1 - \fr{\ml}{\pp} \right),
\nonumber\\
\sin^2 \th_C &\approx& 
\sum_{d} (-1)^{(d+2)/2} 
\epd \ml^{d-4}
\left( 1 - \fr{\ml}{\pp} \right).
\label{sind}
\eea
The latter result reveals a cutoff at large graviton momentum $\vec l$
given by
\beq
\ml_{\rm max} \approx \pp.
\eeq
The maximum momentum of a radiated graviton
is thus approximately the momentum of the incoming particle,
and this provides an upper cutoff 
to the energy-loss integral \rf{rated}.

The explicit form of the integral \rf{rated}
depends on the matrix element for the graviton emission,
but dimensional analysis shows that its basic structure is universal
in the ultrarelativistic limit of interest here.
At tree level,
the emission of a graviton is proportional
to the Newton gravitational constant $\G$.
Also,
since the \cv\ process is forbidden in conventional physics,
the decay rate and the energy-loss rate must be controlled 
by the relevant dimensionless combination $n-1$
of coefficients for Lorentz violation.
The calculations of matrix elements performed below reveal 
that this dimensionless factor is $(n-1)^2$.
Incorporating the various contributions to $n$ from different $d$ 
generates cross terms that complicate the calculation,
so for definiteness and simplicity
we proceed here under the assumption 
that only a single value of $d$ is of interest for a given analysis.
The energy-loss rate then becomes proportional to $(\epd)^2$.
The remaining dimensional factors must involve powers
of the incoming momentum $\pp$.
The result of the integration \rf{rated} therefore takes the form
\beq
\fr{dE}{dt} = - \Cd \G (\epd)^2 \pp^{2d-4} ,
\label{enloss}
\eeq
where $\Cd$ is a dimensionless factor depending on $d$
and on the flavor $w$ of the particle 
emitting the gravitational \cv\ radiation. 
The time of flight of the particle
from its initial energy $E_i$ to a final energy $E_f$ 
is then
\beq
t = \fr {\Fd} {\G (\epd)^2}
\left(\fr{1}{E_f^{2d-5}}-\fr{1}{E_i^{2d-5}}\right),
\label{tflight}
\eeq
where $\Fd \equiv (2d-5)/\Cd$ is another dimensionless factor.

To obtain explicit expressions for $\Cd$ and $\Fd$,
we must consider the matrix elements for specific processes.
Here,
we discuss in turn the cases where the incoming particle
is a scalar, a photon, and a fermion.

\bigskip
\noindent
{\it Scalars radiating gravitons.}
We first consider gravitational \cv\ radiation 
from a hypothetical real massive scalar
minimally coupled to gravity via the Lagrange density
\beq
\cL = - \half e g^\mn \prt_\mu \ph \prt_\nu \ph - \half e m_\ph^2 \ph^2,
\eeq
where $e=\sqrt{|g|}$ is the vierbein determinant.
The Feynman rule for the scalar-scalar-graviton vertex
is $i \kk C^\ph_\mn$,
where
\beq
C^\ph_\mn = - p_\mu k_\nu + \half \et_\mn (p^\al k_\al + m_\ph^2).
\eeq
The second term has no effect in practice as it generates
a trace of the graviton polarization in the matrix element
and therefore vanishes for physical states.
The squared matrix element for the tree-level process takes the form
\beq
|\cM|^2 = \kkt  C^\ph_\mn C^\ph_{\al \be} \ep_r^\mn \ep_r^{\al \be},
\eeq
where $\ep_r^\ab$ with $r=+ ,\times$
are the two physical graviton polarization modes
contained in the matrix \rf{gravpol}.
Summing over these modes and inserting the \cv\ angle,
we find
\beq
|\cM|^2 = 16\pi \G  (\epd)^2 \ml^{2d-8} 
\left(\pp^4 - 2\ml \pp^3 +\pp^2 \ml^2 \right).
\eeq
Upon integration,
the dimension-dependent factor $\Fdw w$ appearing in Eq.\ \rf{tflight} 
is found for massive scalars $w \equiv \ph$ to be
\beq
\Fdw \ph = \frac 1 8 (d-2)(d-3).
\label{cdwscalar}
\eeq
In the special limit with only operators of mass dimension $d=4$
and only isotropic effects,
the results match those obtained in Ref.\ \cite{mo01}. 

\bigskip
\noindent
{\it Photons radiating gravitons.}
Next, 
consider gravitational \cv\ radiation from a high-energy photon.
The electromagnetic part of the Einstein-Maxwell Lagrange density is
\beq
\cL = - \quar e g^{\al \mu} g^{\be \nu} F_\ab F_\mn ,
\label{emlag}
\eeq
where $F_\mn = \prt_\mu A_\nu - \prt_\nu A_\mu$
is the usual electromagnetic field strength.
In TT gauge,
the vierbein determinant contributes only at nonlinear order
in the metric fluctuation and can be neglected here.
Using the standard plane-wave ansatz for the photon,
the Feynman rule for the photon-photon-graviton vertex is found to be
$i\kk C_{\mu \nu \la \rho}$,
where
\bea
C_{\mu \nu \la \rho} = 
- \half \eta_{\mu \nu} (p_\la k_\rho + k_\la p_\rho)
+ \half p_\nu ( \eta_{\la \mu} k_\rho + \eta_{\rho \mu} k_\la)
\nonumber\\
+ \half k_\mu ( \eta_{\la \nu} p_\rho + \eta_{\rho \nu} p_\la)
- \half p_\alpha k_\alpha 
(\eta_{\la \mu} \eta_{\rho \nu} + \eta_{\la \nu} \eta_{\rho \mu}).
\eea
The square of the matrix element for the tree-level diagram is then
\beq
|\cM|^2 = 2\pi\G C_{\mu \nu \la \rho} C_{\al \be \ga \de} 
\ep^\mu_s \ep^\al_s \ep^{\prime \nu}_t \ep^{\prime \be}_t 
\ep^{\la \rh}_r \ep^{\ga \de}_r,
\label{ph_matrix1}
\eeq
where the physical photon polarization vectors are $\ep^\mu_s$
with $s=1,2$
and repeated indices $r$, $s$, and $t$ indicate sums over polarizations.
An extra factor of 1/2 has been incorporated as usual
for the sum over incoming photon polarizations.

In evaluating the result \rf{ph_matrix1},
the condition 
$p^\mu k^\nu C_{\mu \nu \la \rho} = 0$
effectively implements the replacement 
$\ep^\mu_r \ep^\al_r \rightarrow \et^{\mu \al}$.
It is convenient to choose the 3 axis along $\vec l$ 
and the 2 axis along the component of $\vec p$ perpendicular to $\vec l$,
which gives
\bea
p_1 = 0,
\quad
p_2 = - p_0 \sin \th,
\quad
p_\mu l^\mu = -p_0 l_0 ( 1 - n \cos \th),
\eea
and leads to the identities 
\bea
p_\mu p_\nu (\ep^+)^{\mu \al} (\ep^+)^{\nu}_{\pt{\nu} \al} =
p_\mu p_\nu (\ep^\times)^{\mu \al} (\ep^\times)^{\nu}_{\pt{\nu} \al} =
\pp^2 \sin^2 \th,
\nonumber\\
(\ep^+)_\mn (\ep^+)^\mn = (\ep^\times)_\mn (\ep^\times)^\mn = 2,
\nonumber\\
p_\mu p_\nu (\ep^+)^\mn = - \pp^2 \sin^2 \th,
\quad
p_\mu p_\nu (\ep^\times)^\mn = 0.
\eea
Using these results simplifies the squared matrix element
to the form
\bea
|\cM|^2 =
8\pi\G (\epd)^2 
\ml^{2d-8}
\hskip 80pt
\nonumber\\
\times
\left( \ml^4 -4 \ml^3 \pp + 3 \ml^2 \pp^2 + 2 \ml \pp^3-\pp^4 \right).
\eea
Integration reveals that the dimension-dependent factor $\Fd$ 
in Eq.\ \rf{tflight} is
\beq
\Fdw \ga  = \fr
{(d-1) (d-2) (d-3) (2d-3) }
{4(4 d^4 - 28 d^3 + 65 d^2 - 62 d + 27) }
\label{cdwphoton}
\eeq
for photons $w \equiv \ga$ radiating gravitons.
Note that a key difference between
the massive-particle and photon cases
is that photons are always above threshold
for gravitational \cv\ radiation.

\bigskip
\noindent
{\it Fermions radiating gravitons.}
With the notation and conventions used in Eq.\ (12) of Ref.\ \cite{akgrav},
the Lagrange density describing the minimal gravitational coupling
of a relativistic fermion can be written as 
\beq
\cL = \half i e e^\mu{}_a \psb \ga^a D_\mu \ps + {\rm ~ h.c.},
\eeq
where the gravitational degrees of freedom
appear in the vierbein $e_\mu{}^a$ 
and in the covariant derivative $D_\mu$ via the spin connection.
The leading contribution to the fermion-fermion-graviton vertex
can be obtained by expanding the vierbein
and noting that the spin connection contributes only at higher order.
The Feynman rule for the vertex takes the form $i \kk C^\ps_\mn$,
where
\beq
C^\ps_\mn = \quar \ga_\mu(p_\nu + k_\nu).
\eeq
The squared matrix element then becomes 
\bea
|\cM|^2 = 8\pi \G \sum_{\rm spins} 
\left( \ol u(p) C^\ps_\mn u(k) \ep^\mn_r \right)
\left( \ol u(k) C^\ps_{\al \be} u(p) \ep^{\al \be}_r \right),
\eea
where a factor of 1/2 is included for the usual average
over the incoming fermion spins.
Using the methodology developed for the photon case
along with the standard projection 
\beq
\sum_{\rm spins} u(p) \otimes \ol u(p) = \ga \cdot p
\eeq
for a relativistic fermion,
we find 
\bea
|\cM|^2 = 8 \pi \G (\epd)^2 \pp \ml^{2d-8} 
\hskip 80pt
\nonumber\\
\times
\left( -\ml^3 +3 \pp \ml^2 - 4 \pp^2 \ml + 2 \pp^3 \right).
\eea
The dimension-dependent factor $\Fd$
in Eq.\ \rf{tflight}
resulting from the integration for $w \equiv \ps$ is
\beq
\Fdw \ps = 
\fr {(d-2)(d-3)(2d-3)} {4(2 d^2 - 7d +9)}. 
\label{cdwfermion}
\eeq
Note that this differs from the scalar result \rf{cdwscalar}
due to an additional term arising from the sum over spins.

\bigskip
\noindent
{\it 4.\ Constraints.}
According to the above results,
the observation of a cosmic ray of species $w$ 
arriving at the Earth with energy $E_f$
after travelling a distance $L$ along the direction $\hat p_\mu$
implies that the coefficients for Lorentz violation
must satisfy the one-sided constraint 
\beq
\hskip -20pt
\epd(\hat p) \equiv (\sb^{(d)})^{\mn\al_1 \ldots \al_{d-4}}
\hat p_\mu \hat p_\nu 
\hat p_{\al_1} \ldots \hat p_{\al_{d-4}}
< \sqrt{\fr{\Fd}{\G E_f^{2d-5}L}}.
\label{bound}
\eeq
This reveals that high-energy particles originating at large distances
offer the best bounds. 

The species dependence in the bound \rf{bound}
is encoded entirely in the factor $\Fd$, 
which is given for scalars, photons, and fermions
in Eqs.\ \rf{cdwscalar}, \rf{cdwphoton}, and \rf{cdwfermion},
respectively.
Note that these factors are finite for any finite $d$,
and they vanish only for physically irrelevant values $d<4$. 
Note also that they imply enhancements in radiated power
for increasing particle spin.
As a result,
for fixed values of $E_f$ and $L$,
photons yield more sensitive bounds than fermions,
and fermions more sensitive ones than scalars. 
The improved sensitivity increases with increasing $d$.
For example,
the ratio $R \equiv \Fdw \ga:\Fdw \ps:\Fdw \ph$ 
is $R\simeq 0.6:0.8:1$ for $d=4$,
but changes to $R\simeq 0.05:0.5:1$ for $d=6$
and becomes $R\to d^{-2}:d^{-1}/2:1$ at large $d$.

To gain some initial intuition 
about the implications of the bound \rf{bound},
consider the conservative scenario
of a heavy nucleus travelling a distance 
$L\simeq 10$ Mpc $\simeq 10^{39}$ GeV$^{-1}$
from a nearby active galactic nucleus
and impacting the Earth
with an observed cosmic-ray energy 
of about 100 EeV.
Assuming the gravitational \cv\ radiation
occurs from a partonic fermion in the nucleus
carrying about $10^{8}$ GeV of the total cosmic-ray energy
and taking the factor $\Fdw \ps$ to be of order 1 for simplicity,
we find constraints on combinations 
of coefficients for Lorentz violation of dimension $4-d$
of approximate order $10^{20-8d}$ GeV$^{4-d}$.
Although only a crude estimate,
this serves to reveal the quality 
of constraints from gravitational \cv\ radiation.
For example,
bounds on some $d=4$ Lorentz-violating operators 
should exceed by several orders of magnitude 
the various existing sensitivities,
which are of order $10^{-5}$-$10^{-10}$
on dimensionless coefficients in the gravity sector
\cite{2007Battat,2007MullerInterf,2009Chung,%
2010Bennett,2012Iorio,2013Bailey,2014Shao,he15,bk}.
Similarly,
limits for the case $d=6$ should reach $10^{-28}$ GeV$^{-2}$ or so,
representing stringent first constraints on 
this class of nonminimal coefficients in the gravity sector. 

Repeating the above crude estimate but 
replacing the impinging cosmic ray with a high-energy photon 
reveals that gravitational \cv\ radiation from photons
generically provides weaker constraints.
For example,
even the observation of an ultra-high-energy gamma ray at 100 TeV
would imply a sensitivity 
to Lorentz-violating operators of dimension $d$
reduced by a factor of $10^{6d-15}$ compared to cosmic rays.
This factor overwhelms any possible gain in sensitivity 
from greater photon propagation distances $L$,
even for the most favorable case with $d=4$
and for photons originating at cosmological distances.
We therefore focus here on constraints
from gravitational \cv\ radiation by cosmic rays,
deferring further consideration of high-energy photons 
to the discussion section below.

\begin{table}
\begin{center}
\begin{tabular}{l|c|c|c}
Observatory & events & $E_{\rm max}$ (EeV) & Ref. \\
\hline
\hline
AGASA	&	22	&	213	&	\cite{cat,agasa}	\\
Fly's Eye	&	1	&	320	&	\cite{high}	\\
Haverah Park	&	13	&	159	&	\cite{cat,hp}	\\
HiRes	&	11	&	127	&	\cite{hires}	\\
Pierre Auger	&	136	&	127	&	\cite{pa}	\\
SUGAR	&	31	&	197	&	\cite{cat,sugar}	\\
Telescope Array	&	60	&	162	&	\cite{ta}	\\
Volcano Ranch	&	2	&	139	&	\cite{cat,vr}	\\
Yakutsk	&	23	&	160	&	\cite{cat,yak}	\\
\end{tabular}
\vskip -5pt
\caption{\label{observatories}
Cosmic-ray events and maximum energies used in this work.} 
\end{center}
\vskip -5pt
\end{table}

\begin{table}
\begin{center}
\begin{tabular}{cc||c|c|c}
$d$ & $j$ & Lower bound & Coefficient & Upper bound \\
\hline
\hline
4	 & 	0	 & 	   $   	-3	\times 10^{  	-14	  }  < $ 	 & 	 $      \cof{  4   }{  0   0   } $ 	 & 	$				  $ 	\\
\hline																	
4	 & 	1	 & 	   $   	-1	\times 10^{  	-13	  }  < $ 	 & 	 $      \cof{  4   }{  1   0   } $ 	 & 	$ <	7	\times 10^{  	-14	  } $   	\\
       	 & 		 & 	   $   	-8	\times 10^{	-14	  }  < $ 	 & 	 $  \re \cof{  4   }{  1   1   } $ 	 & 	$ <	8	\times 10^{  	-14	  } $   	\\
       	 & 	       	 & 	   $   	-7	\times 10^{	-14	  }  < $ 	 & 	 $  \im \cof{  4   }{  1   1   } $ 	 & 	$ <	9	\times 10^{  	-14	  } $   	\\
\hline																	
4	 & 	2	 & 	   $   	-7	\times 10^{	-14	  }  < $ 	 & 	 $      \cof{  4   }{  2   0   } $ 	 & 	$ <	1	\times 10^{  	-13	  }  $   	\\
       	 & 	       	 & 	   $   	-7	\times 10^{	-14	  }  < $ 	 & 	 $  \re \cof{  4   }{  2   1   } $ 	 & 	$ <	7	\times 10^{  	-14	  }  $   	\\
       	 & 	       	 & 	   $   	-5	\times 10^{	-14	  }  < $ 	 & 	 $  \im \cof{  4   }{  2   1   } $ 	 & 	$ <	8	\times 10^{  	-14	  }  $   	\\
       	 & 	       	 & 	   $   	-6	\times 10^{	-14	  }  < $ 	 & 	 $  \re \cof{  4   }{  2   2   } $ 	 & 	$ <	8	\times 10^{  	-14	  }  $   	\\
       	 & 	       	 & 	   $   	-7	\times 10^{	-14	  }  < $ 	 & 	 $  \im \cof{  4   }{  2   2   } $ 	 & 	$ <	7	\times 10^{  	-14	  }  $   	\\
\end{tabular}
\vskip -5pt
\caption{\label{d4}
Conservative constraints on dimensionless coefficients $\cof{4}{jm}$.} 
\end{center}
\vskip -10pt
\end{table}

\begin{table}
\begin{center}
\begin{tabular}{cc||c|c|c}
$d$ & $j$ & Lower bound & Coefficient & Upper bound \\
\hline\hline
6	 & 	0	 & 	   $   				  $ 	 & 	 $      \cof{  6   }{  0   0   } $ 	 & 	$ <	2	\times 10^{  	-31	  }  $	\\
\hline																	
6	 & 	1	 & 	   $   	-6	\times 10^{	-30	  }  < $ 	 & 	 $      \cof{  6   }{  1   0   } $ 	 & 	$ <	1	\times 10^{  	-29	  }  $   	\\
       	 & 	       	 & 	   $   	-6	\times 10^{	-30	  }  < $ 	 & 	 $  \re \cof{  6   }{  1   1   } $ 	 & 	$ <	7	\times 10^{  	-30	  }  $   	\\
       	 & 	       	 & 	   $   	-8	\times 10^{	-30	  }  < $ 	 & 	 $  \im \cof{  6   }{  1   1   } $ 	 & 	$ <	5	\times 10^{  	-30	  }  $   	\\
\hline																	
6	 & 	2	 & 	   $   	-1	\times 10^{	-29	  }  < $ 	 & 	 $      \cof{  6   }{  2   0   } $ 	 & 	$ <	1	\times 10^{  	-29	  }  $   	\\
       	 & 	       	 & 	   $   	-7	\times 10^{	-30	  }  < $ 	 & 	 $  \re \cof{  6   }{  2   1   } $ 	 & 	$ <	7	\times 10^{  	-30	  }  $   	\\
       	 & 	       	 & 	   $   	-9	\times 10^{	-30	  }  < $ 	 & 	 $  \im \cof{  6   }{  2   1   } $ 	 & 	$ <	6	\times 10^{  	-30	  }  $   	\\
       	 & 	       	 & 	   $   	-9	\times 10^{	-30	  }  < $ 	 & 	 $  \re \cof{  6   }{  2   2   } $ 	 & 	$ <	6	\times 10^{  	-30	  }  $   	\\
       	 & 	       	 & 	   $   	-8	\times 10^{	-30	  }  < $ 	 & 	 $  \im \cof{  6   }{  2   2   } $ 	 & 	$ <	6	\times 10^{  	-30	  }  $   	\\
\hline																	
6	 & 	3	 & 	   $   	-1	\times 10^{	-29	  }  < $ 	 & 	 $      \cof{  6   }{  3   0   } $ 	 & 	$ <	8	\times 10^{  	-30	  }  $   	\\
       	 & 	       	 & 	   $   	-8	\times 10^{	-30	  }  < $ 	 & 	 $  \re \cof{  6   }{  3   1   } $ 	 & 	$ <	7	\times 10^{  	-30	  }  $   	\\
       	 & 	       	 & 	   $   	-6	\times 10^{	-30	  }  < $ 	 & 	 $  \im \cof{  6   }{  3   1   } $ 	 & 	$ <	6	\times 10^{  	-30	  }  $   	\\
       	 & 	       	 & 	   $   	-6	\times 10^{	-30	  }  < $ 	 & 	 $  \re \cof{  6   }{  3   2   } $ 	 & 	$ <	6	\times 10^{  	-30	  }  $   	\\
       	 & 	       	 & 	   $   	-7	\times 10^{	-30	  }  < $ 	 & 	 $  \im \cof{  6   }{  3   2   } $ 	 & 	$ <	7	\times 10^{  	-30	  }  $   	\\
       	 & 	       	 & 	   $   	-7	\times 10^{	-30	  }  < $ 	 & 	 $  \re \cof{  6   }{  3   3   } $ 	 & 	$ <	5	\times 10^{  	-30	  }  $   	\\
       	 & 	       	 & 	   $   	-7	\times 10^{	-30	  }  < $ 	 & 	 $  \im \cof{  6   }{  3   3   } $ 	 & 	$ <	8	\times 10^{  	-30	  }  $   	\\
\hline	   		   						   		   						
6	 & 	4	 & 	   $   	-1	\times 10^{	-29	  }  < $ 	 & 	 $      \cof{  6   }{  4   0   } $ 	 & 	$ <	7	\times 10^{  	-30	  }  $   	\\
       	 & 	       	 & 	   $   	-8	\times 10^{	-30	  }  < $ 	 & 	 $  \re \cof{  6   }{  4   1   } $ 	 & 	$ <	5	\times 10^{  	-30	  }  $   	\\
       	 & 	       	 & 	   $   	-8	\times 10^{	-30	  }  < $ 	 & 	 $  \im \cof{  6   }{  4   1   } $ 	 & 	$ <	7	\times 10^{  	-30	  }  $   	\\
       	 & 	       	 & 	   $   	-7	\times 10^{	-30	  }  < $ 	 & 	 $  \re \cof{  6   }{  4   2   } $ 	 & 	$ <	6	\times 10^{  	-30	  }  $   	\\
       	 & 	       	 & 	   $   	-8	\times 10^{	-30	  }  < $ 	 & 	 $  \im \cof{  6   }{  4   2   } $ 	 & 	$ <	6	\times 10^{	-30	  }  $   	\\
       	 & 	       	 & 	   $   	-9	\times 10^{	-30	  }  < $ 	 & 	 $  \re \cof{  6   }{  4   3   } $ 	 & 	$ <	4	\times 10^{	-30	  }  $   	\\
       	 & 	       	 & 	   $   	-6	\times 10^{	-30	  }  < $ 	 & 	 $  \im \cof{  6   }{  4   3   } $ 	 & 	$ <	7	\times 10^{	-30	  }  $   	\\
       	 & 	       	 & 	   $   	-8	\times 10^{	-30	  }  < $ 	 & 	 $  \re \cof{  6   }{  4   4   } $ 	 & 	$ <	9	\times 10^{	-30	  }  $   	\\
       	 & 	       	 & 	   $   	-4	\times 10^{	-30	  }  < $ 	 & 	 $  \im \cof{  6   }{  4   4   } $ 	 & 	$ <	8	\times 10^{	-30	  }  $   	\\
\end{tabular}
\vskip -5pt
\caption{\label{d6}
Conservative constraints on coefficients $\cof{6}{jm}$ in GeV$^{-2}$.}
\end{center}
\vskip -10pt
\end{table}

To obtain more definitive constraints,
information about the direction of travel of the cosmic rays
is required,
in addition to their energy and distance of travel.
Since cosmic rays impinge upon the Earth from many directions
on the celestial sphere,
it is natural to work with coefficients for Lorentz violation
expressed in spherical coordinates
rather than cartesian ones
\cite{km09}. 
The combination $\epd$ of coefficients,
which appears in the refractive index \rf{nepd} 
and is given in terms of coefficients for Lorentz violation
by Eq.\ \rf{refinds},
is an observer scalar and hence can be expanded
in terms of spherical harmonics as
\beq
\epd(\hat p) = \sum_{jm} Y_{jm}(\hat p) 
\cof{d}{jm},
\eeq
where $jm$ are the usual angular quantum numbers,
subject here to the restriction that $j$ is even and $j\leq d-2$.
At each fixed $d$,
there are $(d-1)^2$ independent spherical coefficients.
The result \rf{bound} then becomes a constraint
on the spherical coefficients $\cof{d}{jm}$.

To obtain conservative bounds,
we suppose that the cosmic ray primary is a nucleus
of atomic weight $N$.
Most high-energy cosmic rays are believed to be protons ($N=1$),
but some may be light nuclei or 
even heavy nuclei such as iron ($N=56)$
\cite{gap}.
We also assume that any gravitational radiation 
is emitted by one of the fermionic partons in the nucleus,
which carries a fraction $r$ of the cosmic-ray energy $E_\oplus$
observed at the Earth.
A conservative estimate is $r = 10$\%
\cite{mo01}.  
We therefore take the energy $E_f$ as 
$E_f = r E_\oplus/N\approx E_\oplus/560$,
with the factor $\Fdw w$ in the bound \rf{bound} 
identified as $\Fdw \ps$ in Eq.\ \rf{cdwfermion}.
This conservative estimate therefore 
represents a reduction by a factor $(N/r)^{d-5/2}\simeq 560^{d-5/2}$
of the effective energy and hence of the bounds.
The acceleration sites of cosmic rays
are believed to be extragalactic,
including possibly supermassive black holes in active galactic nuclei
\cite{hkpt}.
The nearest of these lies at a distance of a few Mpc,
which offers a sense of the minimum value 
of the distance $L$.
The distance is limited by spallation of the cosmic ray 
on photons in the cosmic microwave background
\cite{gzk}.
Approximately 50\% of protons and iron nuclei 
are believed to survive at distances of 100 Mpc,
while for lighter nuclei the analogous distance is 20 Mpc
\cite{kot11}.
For definiteness,
we take $L \simeq 10$ Mpc.

Numerous collaborations have published data for the energies
and angular positions of observed cosmic rays.
Since higher-energy events provide greater sensitivity
to coefficients for Lorentz violation,
we restrict attention here to events with energies above 60 EeV.
Table \ref{observatories} provides some information
about 299 observed events of this type.
The first column lists the cosmic-ray observatory,
the second shows the number of published events above 60 EeV
used in this analysis,
the third gives the maximum observed energy $E_{\rm max}$,
and the final column provides the reference. 
To obtain numerical constraints,
we adopt the modified simplex method of linear programming 
\cite{lp}
detailed in Ref.\ \cite{di14}
in the context of bounds on Lorentz violation 
from nongravitational \cv\ radiation by neutrinos. 
In the present instance,
the available dataset of 299 events is sufficiently large
to place constraints on all coefficients
for fixed $d=4$, $d=6$, and $d=8$ in turn.
In principle,
higher values of $d$ could also be considered,
and additional cosmic-ray data for energies below 60 EeV
could be included as well.

\begin{table}
\begin{center}
\begin{tabular}{cc||c|c|c}
$d$ & $j$ & Lower bound & Coefficient & Upper bound \\
\hline\hline
8	 & 	0	 & 	   $   	-7	\times 10^{  	-49	  }  < $ 	 & 	 $      \cof{  8   }{  0   0   } $ 	 & 	$				  $ 	\\
\hline																	
8	 & 	1	 & 	   $   	-1	\times 10^{	-45	  }  < $ 	 & 	 $      \cof{  8   }{  1   0   } $ 	 & 	$ <	1	\times 10^{  	-45	  }  $   	\\
       	 & 	       	 & 	   $   	-9	\times 10^{	-46	  }  < $ 	 & 	 $  \re \cof{  8   }{  1   1   } $ 	 & 	$ <	8	\times 10^{  	-46	  }  $   	\\
       	 & 	       	 & 	   $   	-9	\times 10^{	-46	  }  < $ 	 & 	 $  \im \cof{  8   }{  1   1   } $ 	 & 	$ <	9	\times 10^{  	-46	  }  $   	\\
\hline																	
8	 & 	2	 & 	   $   	-9	\times 10^{	-46	  }  < $ 	 & 	 $      \cof{  8   }{  2   0   } $ 	 & 	$ <	1	\times 10^{  	-45	  }  $   	\\
       	 & 	       	 & 	   $   	-1	\times 10^{	-45	  }  < $ 	 & 	 $  \re \cof{  8   }{  2   1   } $ 	 & 	$ <	8	\times 10^{  	-46	  }  $   	\\
       	 & 	       	 & 	   $   	-8	\times 10^{	-46	  }  < $ 	 & 	 $  \im \cof{  8   }{  2   1   } $ 	 & 	$ <	9	\times 10^{  	-46	  }  $   	\\
       	 & 	       	 & 	   $   	-1	\times 10^{	-45	  }  < $ 	 & 	 $  \re \cof{  8   }{  2   2   } $ 	 & 	$ <	9	\times 10^{  	-46	  }  $   	\\
       	 & 	       	 & 	   $   	-1	\times 10^{	-45	  }  < $ 	 & 	 $  \im \cof{  8   }{  2   2   } $ 	 & 	$ <	9	\times 10^{  	-46	  }  $   	\\
\hline																	
8	 & 	3	 & 	   $   	-1	\times 10^{	-45	  }  < $ 	 & 	 $      \cof{  8   }{  3   0   } $ 	 & 	$ <	1	\times 10^{  	-45	  }  $   	\\
       	 & 	       	 & 	   $   	-1	\times 10^{	-45	  }  < $ 	 & 	 $  \re \cof{  8   }{  3   1   } $ 	 & 	$ <	8	\times 10^{  	-46	  }  $   	\\
       	 & 	       	 & 	   $   	-9	\times 10^{	-46	  }  < $ 	 & 	 $  \im \cof{  8   }{  3   1   } $ 	 & 	$ <	9	\times 10^{  	-46	  }  $   	\\
       	 & 	       	 & 	   $   	-8	\times 10^{	-46	  }  < $ 	 & 	 $  \re \cof{  8   }{  3   2   } $ 	 & 	$ <	9	\times 10^{  	-46	  }  $   	\\
       	 & 	       	 & 	   $   	-9	\times 10^{	-46	  }  < $ 	 & 	 $  \im \cof{  8   }{  3   2   } $ 	 & 	$ <	8	\times 10^{  	-46	  }  $   	\\
       	 & 	       	 & 	   $   	-8	\times 10^{	-46	  }  < $ 	 & 	 $  \re \cof{  8   }{  3   3   } $ 	 & 	$ <	1	\times 10^{  	-45	  }  $   	\\
       	 & 	       	 & 	   $   	-1	\times 10^{	-45	  }  < $ 	 & 	 $  \im \cof{  8   }{  3   3   } $ 	 & 	$ <	1	\times 10^{  	-45	  }  $   	\\
\hline	   		   						   		   						
8	 & 	4	 & 	   $   	-1	\times 10^{	-45	  }  < $ 	 & 	 $      \cof{  8   }{  4   0   } $ 	 & 	$ <	1	\times 10^{  	-45	  }  $   	\\
       	 & 	       	 & 	   $   	-6	\times 10^{	-46	  }  < $ 	 & 	 $  \re \cof{  8   }{  4   1   } $ 	 & 	$ <	1	\times 10^{  	-45	  }  $   	\\
       	 & 	       	 & 	   $   	-8	\times 10^{	-46	  }  < $ 	 & 	 $  \im \cof{  8   }{  4   1   } $ 	 & 	$ <	1	\times 10^{  	-45	  }  $   	\\
       	 & 	       	 & 	   $   	-8	\times 10^{	-46	  }  < $ 	 & 	 $  \re \cof{  8   }{  4   2   } $ 	 & 	$ <	1	\times 10^{  	-45	  }  $   	\\
       	 & 	       	 & 	   $   	-6	\times 10^{	-46	  }  < $ 	 & 	 $  \im \cof{  8   }{  4   2   } $ 	 & 	$ <	1	\times 10^{	-45	  }  $   	\\
       	 & 	       	 & 	   $   	-7	\times 10^{	-46	  }  < $ 	 & 	 $  \re \cof{  8   }{  4   3   } $ 	 & 	$ <	1	\times 10^{	-45	  }  $   	\\
       	 & 	       	 & 	   $   	-8	\times 10^{	-46	  }  < $ 	 & 	 $  \im \cof{  8   }{  4   3   } $ 	 & 	$ <	8	\times 10^{	-46	  }  $   	\\
       	 & 	       	 & 	   $   	-1	\times 10^{	-45	  }  < $ 	 & 	 $  \re \cof{  8   }{  4   4   } $ 	 & 	$ <	8	\times 10^{	-46	  }  $   	\\
       	 & 	       	 & 	   $   	-9	\times 10^{	-46	  }  < $ 	 & 	 $  \im \cof{  8   }{  4   4   } $ 	 & 	$ <	6	\times 10^{	-46	  }  $   	\\
\hline	   		   						   		   						
8	 & 	5	 & 	   $   	-1	\times 10^{	-45	  }  < $ 	 & 	 $      \cof{  8   }{  5   0   } $ 	 & 	$ <	1	\times 10^{  	-45	  }  $   	\\
       	 & 	       	 & 	   $   	-8	\times 10^{	-46	  }  < $ 	 & 	 $  \re \cof{  8   }{  5   1   } $ 	 & 	$ <	1	\times 10^{  	-45	  }  $   	\\
       	 & 	       	 & 	   $   	-8	\times 10^{	-46	  }  < $ 	 & 	 $  \im \cof{  8   }{  5   1   } $ 	 & 	$ <	7	\times 10^{  	-46	  }  $   	\\
       	 & 	       	 & 	   $   	-9	\times 10^{	-46	  }  < $ 	 & 	 $  \re \cof{  8   }{  5   2   } $ 	 & 	$ <	9	\times 10^{  	-46	  }  $   	\\
       	 & 	       	 & 	   $   	-8	\times 10^{	-46	  }  < $ 	 & 	 $  \im \cof{  8   }{  5   2   } $ 	 & 	$ <	8	\times 10^{	-46	  }  $   	\\
       	 & 	       	 & 	   $   	-1	\times 10^{	-45	  }  < $ 	 & 	 $  \re \cof{  8   }{  5   3   } $ 	 & 	$ <	7	\times 10^{	-46	  }  $   	\\
       	 & 	       	 & 	   $   	-6	\times 10^{	-46	  }  < $ 	 & 	 $  \im \cof{  8   }{  5   3   } $ 	 & 	$ <	1	\times 10^{	-45	  }  $   	\\
       	 & 	       	 & 	   $   	-9	\times 10^{	-46	  }  < $ 	 & 	 $  \re \cof{  8   }{  5   4   } $ 	 & 	$ <	1	\times 10^{	-45	  }  $   	\\
       	 & 	       	 & 	   $   	-8	\times 10^{	-46	  }  < $ 	 & 	 $  \im \cof{  8   }{  5   4   } $ 	 & 	$ <	8	\times 10^{	-46	  }  $   	\\
       	 & 	       	 & 	   $   	-8	\times 10^{	-46	  }  < $ 	 & 	 $  \re \cof{  8   }{  5   5   } $ 	 & 	$ <	1	\times 10^{	-45	  }  $   	\\
       	 & 	       	 & 	   $   	-8	\times 10^{	-46	  }  < $ 	 & 	 $  \im \cof{  8   }{  5   5   } $ 	 & 	$ <	1	\times 10^{	-45	  }  $   	\\
\hline	   		   						   		   						
8	 & 	6	 & 	   $   	-1	\times 10^{	-45	  }  < $ 	 & 	 $      \cof{  8   }{  6   0   } $ 	 & 	$ <	2	\times 10^{  	-45	  }  $   	\\
       	 & 	       	 & 	   $   	-8	\times 10^{	-46	  }  < $ 	 & 	 $  \re \cof{  8   }{  6   1   } $ 	 & 	$ <	1	\times 10^{  	-45	  }  $   	\\
       	 & 	       	 & 	   $   	-7	\times 10^{	-46	  }  < $ 	 & 	 $  \im \cof{  8   }{  6   1   } $ 	 & 	$ <	9	\times 10^{  	-46	  }  $   	\\
       	 & 	       	 & 	   $   	-1	\times 10^{	-45	  }  < $ 	 & 	 $  \re \cof{  8   }{  6   2   } $ 	 & 	$ <	6	\times 10^{  	-46	  }  $   	\\
       	 & 	       	 & 	   $   	-6	\times 10^{	-46	  }  < $ 	 & 	 $  \im \cof{  8   }{  6   2   } $ 	 & 	$ <	1	\times 10^{	-45	  }  $   	\\
       	 & 	       	 & 	   $   	-7	\times 10^{	-46	  }  < $ 	 & 	 $  \re \cof{  8   }{  6   3   } $ 	 & 	$ <	1	\times 10^{	-45	  }  $   	\\
       	 & 	       	 & 	   $   	-7	\times 10^{	-46	  }  < $ 	 & 	 $  \im \cof{  8   }{  6   3   } $ 	 & 	$ <	8	\times 10^{	-46	  }  $   	\\
       	 & 	       	 & 	   $   	-8	\times 10^{	-46	  }  < $ 	 & 	 $  \re \cof{  8   }{  6   4   } $ 	 & 	$ <	1	\times 10^{	-45	  }  $   	\\
       	 & 	       	 & 	   $   	-9	\times 10^{	-46	  }  < $ 	 & 	 $  \im \cof{  8   }{  6   4   } $ 	 & 	$ <	8	\times 10^{	-46	  }  $   	\\
       	 & 	       	 & 	   $   	-8	\times 10^{	-46	  }  < $ 	 & 	 $  \re \cof{  8   }{  6   5   } $ 	 & 	$ <	9	\times 10^{	-46	  }  $   	\\
       	 & 	       	 & 	   $   	-8	\times 10^{	-46	  }  < $ 	 & 	 $  \im \cof{  8   }{  6   5   } $ 	 & 	$ <	9	\times 10^{	-46	  }  $   	\\
       	 & 	       	 & 	   $   	-1	\times 10^{	-45	  }  < $ 	 & 	 $  \re \cof{  8   }{  6   6   } $ 	 & 	$ <	9	\times 10^{	-46	  }  $   	\\
       	 & 	       	 & 	   $   	-7	\times 10^{	-46	  }  < $ 	 & 	 $  \im \cof{  8   }{  6   6   } $ 	 & 	$ <	1	\times 10^{	-45	  }  $   	\\
\end{tabular}
\caption{\label{d8}
Conservative constraints on coefficients $\cof{8}{jm}$ in GeV$^{-4}$.}
\end{center}
\end{table}

Although the bound \rf{bound} is one sided for each cosmic ray,
the dependence on the direction of travel
and the plethora of data across much of the celestial sphere
mean that independent two-sided constraints 
are implied for almost all spherical coefficients $\cof{d}{jm}$
at each fixed $d$.
The exception is the isotropic coefficient $\cof{d}{00}$,
which produces orientation-independent effects
and hence can only be constrained on one side.
For definiteness,
we perform two independent analyses at each $d=4,6,8$.
One assumes only the isotropic coefficient $\cof{d}{00}$ is nonzero
and yields a single one-sided constraint.
The second assumes a purely anisotropic model
allowing all the coefficients $\cof{d}{jm}$ with $j\neq 0$
to be simultaneously nonzero
and yields $d(d-2)$ independent two-sided constraints.
Note that the spherical coefficients are complex when $m\neq 0$,
so their real and imaginary parts must be treated as independent
for this analysis.

Tables \ref{d4}, \ref{d6}, and \ref{d8} 
contain the resulting constraints
on the spherical coefficients 
$\cof{4}{jm}$,
$\cof{6}{jm}$,
and $\cof{8}{jm}$,
respectively,
obtained using cosmic-ray data 
and reported in the Sun-centered frame
\cite{sunframe}.
The initial two columns provide the values of $d$ and $j$,
while the corresponding spherical coefficients
are listed in the third column.
The numerical lower and upper bounds 
are given in the second and fourth columns,
respectively,
with units as specified in the table captions. 
For coefficients with $d=4$,
the results in Table \ref{d4}
represent improvements of factors of a thousand
to a billion over existing maximal sensitivities
obtained via direct laboratory measurements 
\cite{tables}.
Note that the connection between the spherical coefficients $\cof{4}{jm}$
and the usual cartesian ones is given by equations
analogous to Eq.\ (130) of Ref.\ \cite{km13}.
For coefficients with $d=6,8$,
the results in Tables \ref{d6} and \ref{d8}
are the first constraints in the literature.
For $d=6$,
they complement the constraints 
on other independent coefficients
obtained in experiments testing short-range gravity
\cite{lo15,hust15,hustiu}.
The reader is reminded that the  constraints in 
tables \ref{d4}, \ref{d6}, and \ref{d8} 
are conservative:
if the cosmic rays are assumed to be protons
instead of iron nuclei
and if the full proton energy is assumed available 
for gravitational \cv\ radiation,
then the displayed bounds for 
$d=4$, 5, and 6 would be sharpened by additional factors of 
more than $10^4$, $10^9$, and $10^{15}$,
respectively,
even for the same propagation distance $L$.

\bigskip
\noindent
{\it 5.\ Discussion.}
In this work,
we have derived properties of gravitational waves
in the presence of a class of Lorentz-violating operators of arbitrary $d$,
used the results to derive energy losses
from gravitational \cv\ radiation,
and performed an analysis of existing cosmic-ray observations
to extract constraints 
on a variety of coefficients for Lorentz violation
with $d=4,6,8$.
With the exception of the bound on the isotropic coefficient $\cof {4} {00}$,
all the measurements reported here are the first constraints
obtained from gravitational \cv\ radiation
on the corresponding Lorentz-violating terms,
and none of the coefficients for $d=6$ or 8 
have previously been constrained in the literature.

The constraints in Tables \ref{d4}, \ref{d6}, and \ref{d8} 
are obtained using cosmic rays rather than photons
because the former are observed at much higher energies
than the latter.
Nonetheless,
the absence of gravitational \cv\ radiation from high-energy photons 
does in principle contain additional information.
Indeed,
in a more general treatment,
each species $w$ would provide distinct constraints
because the particle itself experiences Lorentz violation
that is flavor dependent
\cite{sme}.

To illustrate this,
consider the Lorentz-violating vacuum dispersion relation 
\beq
\nw^2 p_0^2 - \vec p^2 - \mw^2 = 0,
\label{dr}
\eeq
where $\nw= \nw (\vec p)$ is a refractive index for the particle
of species $w$,
with $m_w=0$ if the particle is a photon.
The motion of the particle then follows a geodesic
in a pseudo-Finsler spacetime
\cite{finsler1,finsler2}.
An example would be a refractive index 
for a massive fermion given by 
\cite{km13,kr10}
\beq
\nw^2 = 1 + 2\cbh_w^\ab \hat{p}_\al \hat{p}_\be ,
\label{drmatter}
\eeq
in analogy with the refractive index \rf{efri} for gravity,
where $\cbh_w^\ab$ is a momentum-dependent coefficient
having expansion of the form \rf{sdef}.
For $d=4$ this reduces to a special case 
of the matter sector of the minimal SME
with a single fermion flavor 
\cite{sme},
with $\cbh_w^\ab = \cb_w^\ab$ 
being a constant dimensionless coefficient for Lorentz violation.
A similar result holds for neutrinos
\cite{km12}
and for photons
\cite{km09},
where the corresponding constant dimensionless coefficient
is conventionally denoted by $\kfb^\ab$.
Repeating the analysis of gravitational \cv\ radiation
generates an energy loss given by Eq.\ \rf{rated} as before,
but with the vacuum \cv\ angle $\th_C$ given instead by 
\bea
\cos \th_C &=& 
\fr {\sqrt{\mw^2 + \vec p^2}} {\pp} 
\fr{[\nw(\pp-\ml)]^2}{\nw(\pp)n(\ml)}
\nonumber\\
&&+ 
\fr{\ml}{2\pp}
\left(1-
\fr{[\nw(\pp-\ml)]^2}{[n(\ml)]^2}
\right) 
\nonumber\\
&&+ 
\fr {\mw^2 + \vec p^2} {2 \ml\pp} 
\left(1-
\fr{[\nw(\pp-\ml)]^2} {[\nw(\pp)]^2}
\right) ,
\label{cagen}
\eea
where as before the arguments of the refractive index
are understood to be oriented along $\hat p$.
This expression reveals that  
the presence and rate of gravitational \cv\ radiation depends
on ratios of the refractive indices
for the particle and the gravitational waves.
Depending on the relative magnitudes 
of the coefficients for Lorentz violation
for the particle and the graviton,
gravitational \cv\ radiation may occur
in a given direction of travel
for only one sign of the correction $\epd$ 
to the graviton refractive index,
for either sign,
or not at all.
Note that the final term depends only on the particle refractive index,
contributing only when $\nw$ depends on momentum and hence only for $d>4$,
implying the particle is experiencing nonminimal Lorentz violation.
Note also that the above expression reduces to the result \rf{ca} 
in the limit $\nw\to 1$, 
as expected.

Inserting the gravitational \cv\ angle \rf{cagen}
into the integral \rf{rated} for the energy loss
and limiting attention to fixed $d$ 
must by dimensional arguments produce a result 
of the general form \rf{enloss} 
but now involving a linear combination of the quadratic terms 
$(\epd)^2$, $\epd(1-n_w)$, and $(1- n_w)^2$,
where $1-n_w$ is given in terms of matter coefficients for Lorentz violation 
by an expression analogous to that for $\epd$ in Eq.\ \rf{refinds}.
It follows that a detailed analysis 
of observations of high-energy particles,
including photons and neutrinos,
would yield constraints on distinct combinations 
of coefficients for Lorentz violation.
However,
photons yield weaker constraints than those from cosmic rays
because observed cosmic-ray energies exceed the highest photon energies
by about a millionfold,
which becomes scaled by the power $d-5/2$ in extracting a bound. 
A similar conclusion holds for neutrinos.
Note that any observed but unexplained absence of ultra-high-energy particles 
such as neutrinos or photons
could in principle be attributed to Lorentz-violating vacuum \cv\ radiation,
including the gravitational \cv\ radiation considered here.
A complete analysis along these various lines would be of interest
but lies beyond our present scope. 

The expression \rf{cagen} for $\th_C$ 
also reveals that if the Lorentz violation is minimal,
so that all Lorentz-violating operators have mass dimension $d=4$
and both $n$ and $\nw$ are momentum independent,
then the properties of the gravitational \cv\ radiation are determined
by differences of SME coefficients for Lorentz violation
in the gravity and matter sectors. 
This leads to an interesting relation between distinct measurements,
as follows.
In this scenario, 
the rate of energy loss for a massive fermion 
undergoing gravitational \cv\ radiation
is governed by the combination
$\sb^\ab + 2\cb_w^\ab$,
while that for a radiating photon is governed by
$\sb^\ab - \kfb^\ab$.
Moreover,
due to the freedom to choose coordinates
without changing the physics,
all nongravitational searches for Lorentz violation
involving these species must involve the combination
$2\cb_w^\ab + \kfb^\ab$
\cite{kt}.
As a consequence,
if analyses yield measurements $M_1$ 
of $\sb^\ab + 2\cb_w^\ab$,
$M_2$ 
of $\sb^\ab - \kfb^\ab$,
and $M_3$ 
of $2\cb_w^\ab + \kfb^\ab$,
then the three measurements must satisfy the condition
\beq
M_1 - M_2 - M_3 = 0.
\eeq
A relation of this type,
relevant for searches for CPT violation with neutral-meson oscillations,
has previously been inferred
among SME coefficients in the quark sector
\cite{akvk}.

The components of the coefficients 
$(\sb^{(d)})^{\mn\al_1 \ldots \al_{d-4}}$
constrained in this work
can also be measured in other ways.
For example,
all the corresponding Lorentz-violating operators for $d>4$ are dispersive
and so can in principle be measured using pulse-spread data 
from a terrestrial gravitational-wave observatory
such as the
Advanced Laser Interferometer Gravitational-Wave Observatory (LIGO)
\cite{aligo}
and Advanced Virgo
\cite{avirgo}
or from a space observatory such as the proposed 
Evolved Laser Interferometer Space Antenna (eLISA)
\cite{elisa},
assuming gravitational waves are indeed detected.
The sensitivity of dispersion measurements of $\epd$
depends on the ratio of the observed pulse width 
to the source distance $L$,
which typically leads to weaker constraints
than those from gravitational \cv\ radiation.
However,
dispersion limits are distinct and unique in detail 
because they involve Lorentz violation 
in the electron and photon sectors
by virtue of the laser interactions with the mirrors.
Moreover,
dispersive measurements involve the group velocity
and hence are sensitive to the gradient of the refractive index 
instead of the refractive index itself,
so the corresponding measurements constrain
different combinations of Lorentz-violating operators 
when more than one value of $d$ is incorporated.

Another interesting open issue is the prospects for 
independent two-sided bounds on the isotropic coefficients 
$\cof{d}{00}{}$,
which cannot be obtained via gravitational \cv\ radiation.
Gravitational-wave observatories such as LIGO and eLISA
are uniquely suited to place dispersion limits 
on these coefficients for operators of dimension $d>4$.
However,
no dispersion occurs for the minimal isotropic coefficient
$\cof{4}{00}{}$,
which makes its two-sided measurement challenging.
One possibility in principle would be to compare 
the time of flight of gravitational waves 
to that of light or neutrinos emitted from the same source.
This has the disadvantage of requiring simultaneous observation
using different techniques.
Other options already used to place constraints on $\cof{4}{00}{}$
include analyses of data from Gravity Probe B
\cite{2013Bailey}
and from pulsar timing 
\cite{2014Shao}.
Methods such as the study of orbital decay rates of binary systems
\cite{hyy}
have the potential to provide interesting sensitivities 
to $\cof{4}{00}{}$ as well.

The derivations in this work may also have implications
for other ideas. 
For example,
the presence of Lorentz violation in quantum electrodynamics
can induce triple photon splitting in the vacuum
and hence reduce the frequency of light
as a function of distance travelled,
suggesting the possibility of modifications
to the usual interpretation of the observed cosmological redshift
\cite{ko03}.
A similar possibility is implied by
the expressions \rf{enloss}, \rf{tflight}, and \rf{cdwphoton}
for the energy loss of a photon due to gravitational \cv\ radiation.
These ideas are conceptually akin to `tired-light' models 
\cite{tired},
which are strongly constrained by the direct observation
of time dilation associated with cosmological redshift
\cite{bl08}.
However,
the energy losses \rf{enloss}
here have distinctive frequency dependence,
and in principle they might only be perturbative 
or only affect part of the observed redshifts,
perhaps such as those associated with supernova studies of dark energy.
A complete discussion of these possibilities
would require analysis of gravitational-wave propagation 
in a cosmological background
instead of the static Minkowski background adopted here.
Nonetheless,
the analysis in the present work suffices to provide simple estimates 
of the possible scale of the effects,
as follows.

For an astrophysical source at small redshift $z$
defined in terms of the photon energies as usual by
$z+1 = {E_i}/{E_f}$,
the luminosity distance $L_{\rm L}$ 
can be written as 
$L_{\rm L} \approx  (z + O(z^2))/H_0$,
where 
$H_0 \simeq 1.5 \times 10^{-42}$ GeV
is the Hubble constant
\cite{sw08}.
Directly expressing the time of flight \rf{tflight}
as a distance $L$ in terms of $z$ gives
\beq
L \approx \fr {\Fd} {\G (\epd)^2 E_i^{2d-5}}
\left( (z+1)^{2d-5}-1 \right).
\eeq
Comparing $L_{\rm L}$ and $L$ reveals that 
the potential contribution of gravitational \cv\ radiation
to the observed cosmological redshift 
is primarily governed by the dimensionless ratio 
\beq
R \equiv \fr {\G (\epd)^2 E_i^{2d-5}} {H_0} 
\simeq
10^4 E_i^3 (n-1)^2,
\eeq
where $E_i$ is measured in GeV
and values $R \gsim 1$ represent substantial effects.
This suggests that redshift modifications from gravitational \cv\ radiation 
are negligible for most practical purposes.
For example,
for the optical frequencies $\simeq$100-900 nm
typically studied in the spectra of type-Ia supernovae,
the energy factor $E_i^3$ is of order $10^{-24}$ GeV$^3$ or smaller,
so a value $R \gsim 1$ would require $n-1 \gsim 10^{10}$,
which is well outside the perturbative regime.
Moreover,
the energy dependence implies that the effect
varies by orders of magnitude over an observed spectrum,
which for large values of $n-1$ would distort
spectra beyond observed limits. 
The requirement of perturbative $n-1$ 
evidently restricts substantial redshift effects
to high-energy photons.
However,
it remains conceivable that a detailed analysis
along the above lines
could extract additional constraints on $\epd$
from precision cosmological measurements.

In the above, 
we consider gravitational \cv\ radiation by photons.
However,
Lorentz violation in the pure-gravity sector
can also cause electromagnetic \cv\ radiation by gravitons,
corresponding to graviton decay.
Discussion of this process is lacking in the literature.
The form of the Einstein-Maxwell Lagrange density \rf{emlag}
reveals that at leading order this process involves two-photon emission,
being governed by a photon-photon-graviton-graviton vertex.
The corresponding amplitude is proportional to $\G$,
and hence on dimensional grounds
the power loss of the graviton takes the form
\beq
\fr{dE}{dt} = - F^\prime (d) \G^2 (n-1)^k \ml^{6} ,
\label{enlossg}
\eeq
where $F^\prime(d)$ is a dimensionless factor depending on $d$
arising from integration of the matrix element,
$k$ is the power of the dimensionless combination $n-1$
emerging from the matrix element,
and $\ml$ is the magnitude of the graviton momentum.
This result implies that the frequency spectrum of gravity waves
detected by a gravitational-wave observatory on or near the Earth
is distorted and downshifted by Lorentz violation.
However,
the effect is far below observational levels in practice,
both because the power loss \rf{enlossg} is proportional to $\G^2$
and because the energy of typical gravitational waves 
is expected to be tiny.
For example,
a gravitational wave in the LIGO band at frequency 100 Hz 
originating in our galaxy
experiences a negligible frequency shift $\de\nu\approx 10^{-150}(n-1)^k$ Hz. 
For similar reasons,
graviton \cv\ decay into other particle species
is negligible as well.

The results in this work
complement those obtained in tests 
of short-range gravity
\cite{lo15,hust15,hustiu}
and thereby improve the coverage of sensitivities
to coefficients for Lorentz violation
in the gravity sector.
Exploring the remaining coefficients 
for even $d$ and the coefficients for odd $d$,
all of which are birefringent,
is an interesting open problem for future research.

\medskip

This work was supported in part 
by the United States Department of Energy
under grant number {DE}-SC0010120
and by the Indiana University Center for Spacetime Symmetries.

\end{document}